\documentclass[12pt,preprint]{aastex} 
\usepackage{amsmath}
\usepackage{color}

\newcommand*{\blue}{\textcolor{blue}}
\shortauthors{Kreidberg}

\begin{document}

\title{\texttt{batman}: BAsic Transit Model cAlculatioN in Python} 

\author{Laura Kreidberg\altaffilmark{1,2}}

\email{E-mail: laura.kreidberg@uchicago.edu}

\altaffiltext{1}{Department of Astronomy and Astrophysics, University of Chicago, 5640 S.~Ellis Ave, Chicago, IL 60637, USA}
\altaffiltext{2}{National Science Foundation Graduate Research Fellow}

\begin{abstract}
I introduce \texttt{batman}, a Python package for modeling exoplanet transit and eclipse light curves.  The \texttt{batman} package supports calculation of light curves for any radially symmetric stellar limb darkening law, using a new integration algorithm for models that cannot be quickly calculated analytically.  The code uses C extension modules to speed up model calculation and is parallelized with OpenMP.  For a typical light curve with 100 data points in transit, \texttt{batman} can calculate one million quadratic limb-darkened models in 30 seconds with a single 1.7 GHz Intel Core i5 processor.  The same calculation takes seven minutes using the four-parameter nonlinear limb darkening model (computed to 1 ppm accuracy).  Maximum truncation error for integrated models is an input parameter that can be set as low as 0.001 ppm, ensuring that the community is prepared for the precise transit light curves we anticipate measuring with upcoming facilities.  The \texttt{batman} package is open source and publicly available at \texttt{\blue{\url{https://github.com/lkreidberg/batman}}}.
\end{abstract}


\section{Introduction}
The transit technique has revolutionized the study of exoplanetary systems.  Thanks largely to the \emph{Kepler} mission, thousands of planets have been discovered with this method \citep{rowe15}. These discoveries have yielded transformative constraints on planet occurrence rates over a wide range of planet sizes, orbital periods, and host star properties \citep{borucki11, youdin11, howard12,fressin13, dong13, morton14}. They have also enabled the first estimates of the occurrence of habitable planets \citep{traub12, dressing13, petigura13, kopparapu13, foremanmackey14, dressing15}.  Transit light curves can even reveal planets' atmospheric temperature structure and composition \citep[e.g.][]{seager00, charbonneau02, lecavelier08, sing11, deming13, knutson14a, fraine14, kreidberg15}.  A number of current and planned observational facilities -- including \emph{K2}, \emph{TESS}, \emph{CHEOPS}, \emph{JWST}, and \emph{PLATO} -- will measure precise transit light curves for thousands of exoplanets that will further advance our understanding of planet formation, evolution, and habitability.

Light curve models are a fundamental tool for transiting exoplanet science, but they are not trivial to compute quickly and accurately.  Accurate calculation is challenging because the model must account for the planet's size and position on the sky, as well as stellar limb darkening, which causes the apparent brightness of the stellar disk to decrease from center to edge.  The  stellar intensity profile can be fit with several functional forms, including a linear limb darkening law \citep{schwarzschild06}, quadratic \citep{kopal50}, square-root \citep{diaz-cordoves92}, logarithmic \citep{klinglesmith70}, exponential \citep{claret03}, and four-parameter nonlinear \citep{claret00}.  For some of these profiles, model transit light curves can be calculated analytically \citep{mandel02, gimenez06, abubekerov13}.  Other profiles do not have analytic solutions, and models must be calculated by numeric integration of the stellar intensity over the disk of the planet. In addition, speed is an important consideration because a large number of models must typically be calculated to make a robust estimation of transit parameters and their uncertainties.

A number of codes are available to calculate transit light curves. \cite{mandel02} provide Fortran and IDL routines to compute models for quadratic and nonlinear limb darkening laws. The software packages \texttt{TAP} \citep{gazak12} and \texttt{EXOFAST} \citep{eastman13} include IDL implementations of the \cite{mandel02} algorithm for quadratic limb darkening. \texttt{JKTEBOP} calculates models in Fortran for a broad range of limb darkening laws \citep{southworth04}. \cite{kjurkchieva13} introduce the pure Python code \texttt{TAC-maker}, which performs numeric integration for arbitrary limb darkening profiles. There are also routines available to model simultaneous transits by one or more bodies \citep{kipping11, pal12}.  Most recently, \cite{parvianinen15} released the Python package \texttt{PyTransit}, which implements analytic models from \cite{mandel02} and \cite{gimenez06}.  

In this paper, I introduce the open source Python package \texttt{batman}.  This package is based on code that was used to model high-precision light curves obtained for atmosphere characterization \citep{kreidberg14b, kreidberg14a, kreidberg15, stevenson14c, stevenson14b, stevenson14a, stevenson14d}.  The \texttt{batman} package enables fast computation of transit light curves for any radially symmetric limb darkening law, and currently supports uniform, linear, quadratic, logarithmic, exponential, and nonlinear limb darkening. Light curves for the first three of these are calculated analytically based on the formalism from \cite{mandel02}.  Models for the remaining cases are computed with an efficient new integration scheme, described in \S\,\ref{section:algorithm}. The package also supports secondary eclipse modeling. I discuss \texttt{batman}'s features and performance in \S\,\ref{section:batman} and conclude in \S\,\ref{section:summary}.

\section{Algorithm}
\label{section:algorithm}
To calculate the fraction $\delta$ of stellar flux blocked by a transiting planet, one must integrate the sky-projected intensity of the star ($I$) over the area obscured by the disk of the planet ($S$):
\begin{equation}
\delta = \iint_S{I \mathrm{d}S}
\label{eqn:surfaceint}
\end{equation}
where $I$ is normalized such that the integrated intensity over the stellar disk is unity. This expression is valid for any general stellar surface brightness map; however, it is slow to evaluate numerically because the differential area elements must be small ($\lesssim 10^{-6}$) in order to achieve better than one part per million (ppm) accuracy.

On the other hand, if the stellar intensity profile is radially symmetric, the two-dimensional calculation in Equation\,\ref{eqn:surfaceint} can be reduced to one dimension and sped up greatly with the following algorithm: 
\begin{equation}
\delta = \sum_{i=1}^{n} I\left(\frac{x_i+x_{i-1}}{2}\right) \left[A(x_{i}, r_p, d) - A(x_{i-1}, r_p, d)\right]
\end{equation}
where $x$ is the normalized radial coordinate $0 < x < 1$, $I(x)$ is the 1D stellar intensity profile, $r_p$ is the planetary radius (in units of stellar radii), $d$ is the separation of centers between the star and the planet (in stellar radii), and $A(x, r_p, d)$ is the area of intersection between two circles of radii $x$ and $r_p$, separated by a distance $d$.  The sum is carried out over the range $x_0 = \textrm{MAX}(d - r_p, 0)$ to $x_n = \textrm{MIN}(d + r_p, 1)$. The intersecting area is given by:

\begin{equation}
A(x, r_p, d) = 
\begin{cases}
x^2\cos^{-1}{u} + r_p^2\cos^{-1}{v} - 0.5\sqrt{w}, & r_p -d < x < r_p + d\\
\pi x^2, & x \le r_p - d\\
\pi r_p^2 & x \ge r_p +d \\
\end{cases}
\end{equation}
where
\begin{eqnarray}
u &=& (d^2+x^2-r_p^2)/(2dx)\\
v &=& (d^2 + r_p^2 -x^2)/(2dr_p) \\
w &=& (-d+x+r_p)(d+x-r_p)(d-x+r_p)(d+x+r_p).
\end{eqnarray}

\begin{figure}
\resizebox{\hsize}{!}{\includegraphics{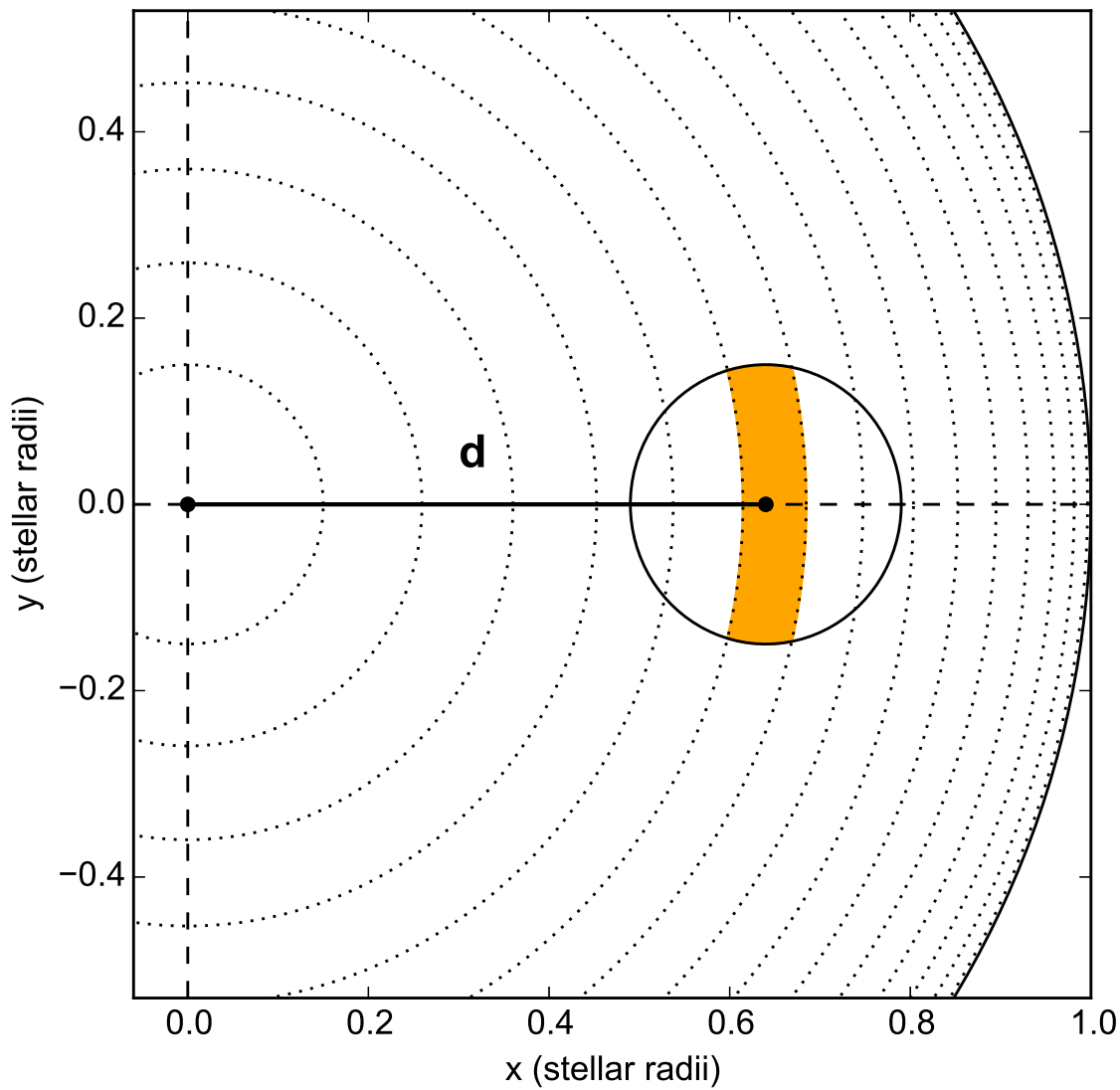}}
\caption{Schematic illustration of the integration scheme. The star (large black circle; partially visible) has a radius of 1 and is centered in the plane of the sky at ($x$, $y$) = (0, 0). The planet (smaller black circle) is separated from the center of the stellar disk by a distance $d$ (marked by the solid black line).  The star is partitioned into concentric circles (dotted lines) in order to calculate the integral over the planet disk.  A single integration element $\Delta A$ is shaded in orange. The integration step size illustrated here is larger than for a typical calculation for visual clarity.  Note that because the stellar intensity profile is radially symmetric, the coordinate system can be chosen such that the planet lies on the x-axis, as shown.}
\label{fig:integration}
\end{figure}

See Figure\,\ref{fig:integration} for a schematic illustrating the geometry of the integration.  The advantage of this integration scheme is that the only error introduced is due to approximating the stellar intensity as constant over the differential area element $\Delta A = A(x_i, r_p, d) - A(x_{i-1}, r_p, d)$.  Computation of $\Delta A$ is expensive, but it needs to be calculated relatively few times ($<<10^{6}$ for sub-ppm accuracy). This makes the integration faster than a scheme with a simpler area element (e.g., $\Delta A = \Delta x \Delta y$), that requires a much smaller step size to achieve the same accuracy.

The integration can be further optimized by using a nonuniform step size.  Typical stellar intensity profiles have larger gradients near the limb of the star than at the center \citep[e.g.][]{claret00}, so smaller steps are required at larger $x$ values to achieve the same accuracy.  I adopt the following step-size scaling: 
$$
x_i - x_{i-1}  = f\cos^{-1}\left(x_{i-1}\right)
$$  
where $f$ is a constant scale factor. This prescription is fast to compute and well-behaved at the limits $x=0$ and $x=1$.

\section{The \texttt{batman} package}
\label{section:batman}
The Python package \texttt{batman} implements the algorithm described in \S\,\ref{section:algorithm} and several analytic models to calculate transit light curves.  \texttt{batman} is an open source project and is being developed on GitHub.  Full documentation is available at \blue{\url{https://github.com/lkreidberg/batman}}. I summarize the main capabilities of the package here.

\subsection{Limb Darkening Models}
\texttt{batman} supports calculation of exoplanet transit light curves for uniform, linear, quadratic, square-root, logarithmic, exponential, and four-parameter nonlinear stellar intensity profiles:
\begin{align}
  I(\mu) &= I_0                                                                         & &\text{(uniform)}             \\
  I(\mu) &= I_0[1 - c_1(1-\mu)]                                                         & &\text{(linear)}              \\
  I(\mu) &= I_0[1 - c_1(1 - \mu) - c_2(1-\mu)^2]                                        & &\text{(quadratic)}           \\
  I(\mu) &= I_0[1 - c_1(1 - \mu) - c_2(1-\sqrt{\mu})]                                   & &\text{(square-root)}         \\
  I(\mu) &= I_0[1 - c_1(1 - \mu) - c_2\mu\ln{\mu}]                                      & &\text{(logarithmic)}         \\
  I(\mu) &= I_0\left[1 - c_1(1 - \mu) - c_2/(1-\exp{\mu})\right]                  	& &\text{(exponential)}         \\
  I(\mu) &= I_0[1 - c_1(1-\mu^{1/2}) - c_2(1- \mu) - c_3(1-\mu^{3/2}) - c_4(1-\mu^2)]   & &\text{(nonlinear)}
\end{align}
where $\mu = \sqrt{1-x^2}$ and $c_1, ..., c_n$ are limb darkening coefficients.  The \texttt{batman} source distribution also includes a template for the creation of a custom profile for any radially symmetric function.  

The square-root, logarithmic, exponential, nonlinear, and custom models are computed with the numeric integration scheme from \S\,\ref{section:algorithm}.  The uniform, linear, and quadratic models are calculated analytically, with code based on the Fortran routines \texttt{occultquad.f} and \texttt{occultuniform.f} provided by \cite{mandel02}. For the analytic models, I follow \cite{eastman13} and use the algorithm from \cite{bulirsch65} to improve calculation speed and accuracy for elliptic integrals of the third kind.

\subsection{Secondary Eclipse Model}
\texttt{batman} can also model secondary eclipses. Eclipse light curves are generated with 
$$
f = 1 + f_p(1-\alpha)
$$
where $f$ is normalized flux, $f_p$ is the planet-to-star flux ratio, and $\alpha$ is the fraction of the planet disk that is occulted by the star. The model is normalized such that the stellar flux is unity.  For a separation $d$, the occultation fraction $\alpha(d) = \alpha_t(d)/r_p^2$, where $1 - \alpha_t(d)$ is the transit light curve with uniform limb darkening. Note that this model assumes the planet flux is constant for all orbital phases.

\subsection{Utilities}
\texttt{batman} includes a utility function to calculate the separation of centers $d$ between the star and the planet based on orbital parameters of the system.  The input parameters are the planet semi-major axis $a$, inclination $i$, eccentricity $e$, longitude of periastron $\omega$, period $P$, and time of inferior conjunction $t_0$.  The separation of centers is given by:
$$
d = \frac{a(1-e^2)}{1+e\cos{f}}\sqrt{1-\sin^2{(\omega+f)}\sin^2{i}}
$$
where $f$ is the true anomaly.  The true anomaly is calculated with the algorithm provided by Murray \& Correia in Chapter 1 of \cite{seager10}. For circular orbits, we adopt the convention $f = \pi/2$.

\texttt{batman} provides utilities to calculate the time of periastron, time of inferior conjunction, and the secondary eclipse time from the other transit parameters, using the method described in \S 3.1 of \cite{eastman13}.  Note however that \texttt{batman} does not correct for the effects of light travel time.

An additional utility is light curve supersampling, which allows the user to calculate the average value of the light curve over a specified number of evenly spaced points during an exposure.

\subsection{Accuracy}
\begin{figure}
\resizebox{\hsize}{!}{\includegraphics{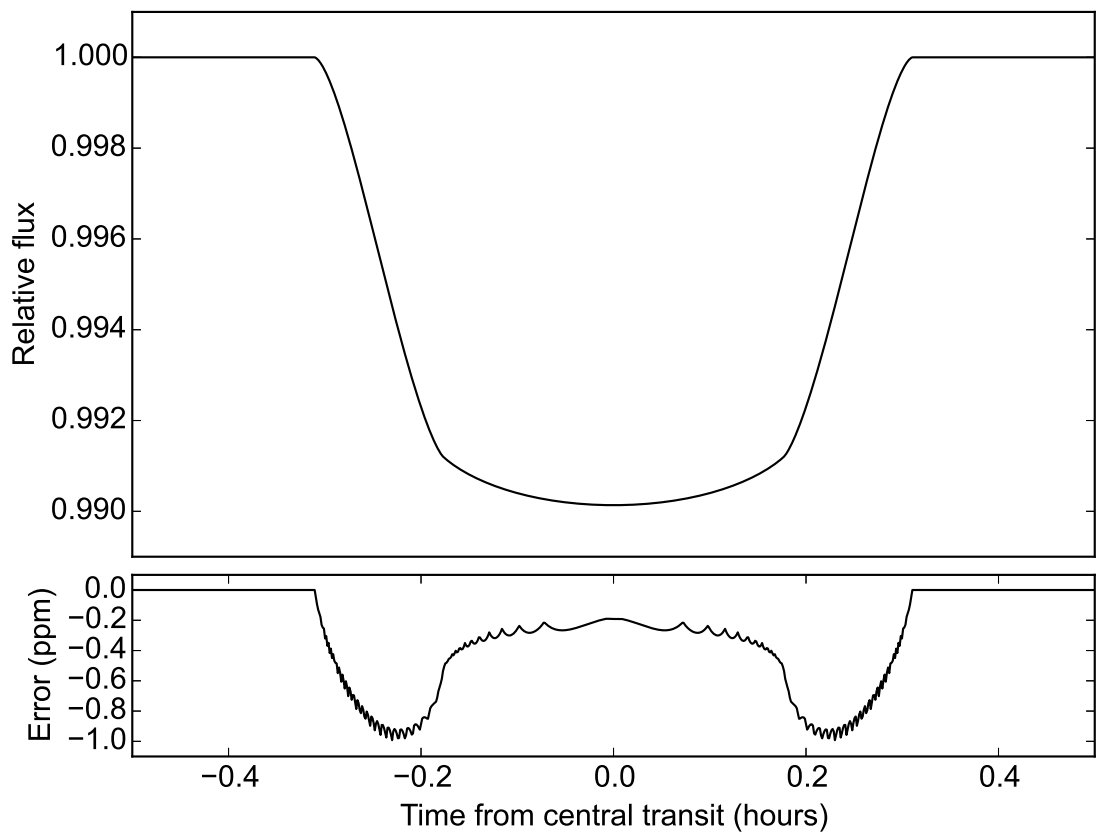}}
\caption{An example transit light curve for a nonlinear stellar intensity profile (top panel) and truncation error for the calculation (bottom panel).  The error tolerance parameter was set to 1.0 ppm.  The truncation error increases with distance from the center of the star up to around $\pm0.2$ hours from the time of mid-transit, because the stellar intensity gradient is larger at larger radii. The error decreases again during ingress and egress as the planet eclipses a smaller fraction of the stellar disk.}
\label{fig:transit}
\end{figure}

Recent transit observations have yielded signal-to-noise greater than 1000 per exposure \citep[e.g.][]{kreidberg14a, knutson14b}.  Accurate transit light curve calculation is essential for modeling such high precision measurements and will be increasingly important for data obtained with next-generation facilities.  \texttt{batman} therefore enables the user to specify the maximum allowable truncation error for numeric integration.  Figure\,\ref{fig:transit} shows an example transit light curve and its truncation error.

To ensure that the truncation error is below the specified threshold, the integration step size is tuned during model initialization.  Truncation error is measured relative to a model calculated with a very small step size ($f = 5\times10^{-4}$).  For typical limb darkening profiles, this method is reliable for truncation errors down to $\sim 10^{-3}$ ppm. However, tuning the step size is a slow operation because it requires computing several light curve models ($\sim10$). As an alternative, methods are available to set the step size directly and calculate the corresponding truncation error. 

I tested the accuracy of the analytic model for quadratic limb darkening by comparing it to a numerically integrated model with an error tolerance of 0.001 ppm.  The analytic model is accurate to 0.03 ppm for a test case with $r_p = 0.1$, $(c_1, c_2) = (0.1, 0.3)$, sampled at $10^6$ evenly spaced points over the interval $0 < d < 1$.  The accuracy is somewhat worse than machine epsilon because of error tolerance in the computation of special functions.

I also tested the accuracy of the widely-used \cite{mandel02} code that uses Numerical Recipes functions to calculate elliptic integrals \citep{press92}.  The accuracy was better than 0.005 ppm for most input values; however, for the case $r_p - d < \epsilon$, the error in the light curve exceeded 2 ppm. By contrast, the \cite{bulirsch65} algorithm for elliptic integrals is well-behaved for this case and also faster.

\subsection{Performance}
\begin{figure}
\resizebox{\hsize}{!}{\includegraphics{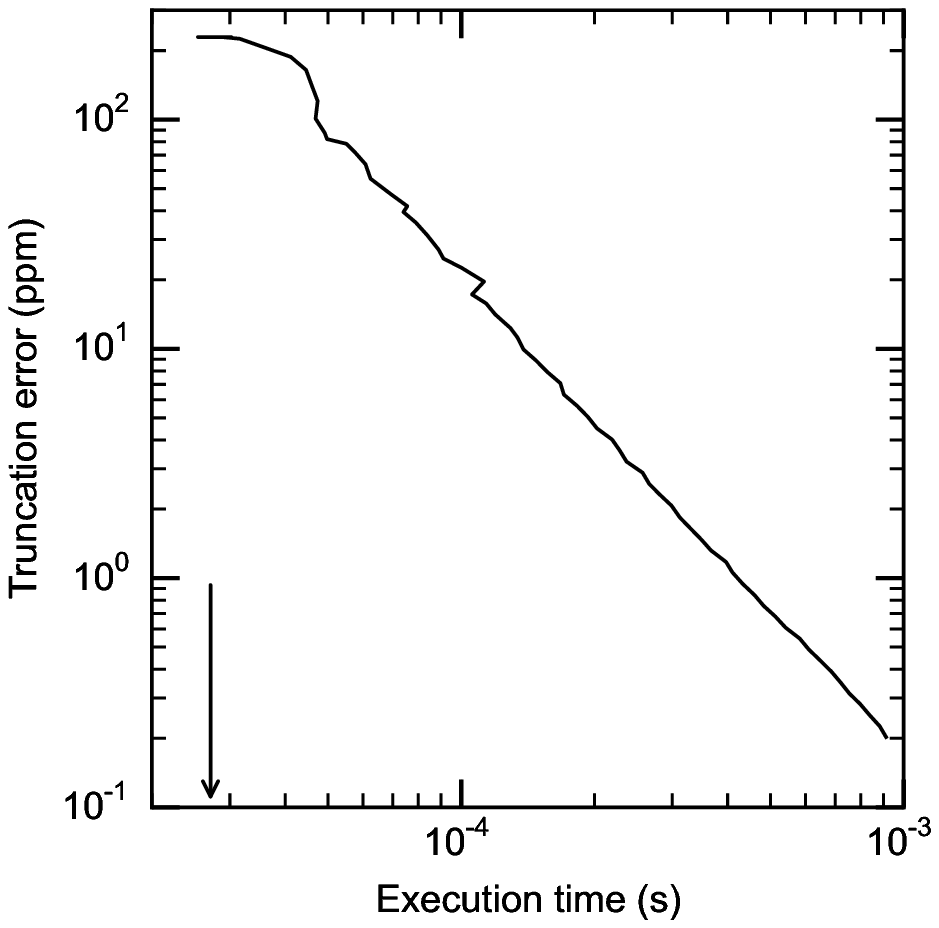}}
\caption{Truncation error as a function of execution time for a light curve modeled with the nonlinear limb darkening law (black line). The execution time for a quadratic model (computed analytically to better than 0.05 ppm accuracy) is indicated by the arrow. Calculations were made with a 1.7 GHz Intel Core i5 processor.} 
\label{fig:performance}
\end{figure}

Computationally intensive sections of code (including all of the transit model calculation) are written as C extension modules with the Python/C API, which improves the performance by a factor of 30 over a pure Python implementation for quadratic limb darkening.  \texttt{batman} also includes the option to parallelize at the C level with OpenMP, which further speeds up the calculation. The number of processors is specified by the user.  \texttt{batman} will raise an exception if the user attempts to parallelize a calculation on a system where OpenMP is not supported.

I tested \texttt{batman}'s performance over a range of typical use cases with a 1.7 GHz Intel Core i5 processor.  In Figure\,\ref{fig:performance}, I show the truncation error versus execution time for a single transit light curve calculation using a nonlinear intensity profile, compared to the execution time for a quadratic model computed analytically.  The test case consists of 100 points evenly sampled in time during the planet's transit. I used physical parameters for the transiting planet GJ\,1214b \citep{kreidberg14a}. The stellar intensity profile is the same for the nonlinear and quadratic models: the nonlinear limb darkening coefficients are $(0.0, 0.7, 0.0, -0.3)$ and the quadratic coefficients are $(0.1, 0.3)$.  Decreasing the truncation error by a factor of 10 increases the computation time by a factor of three.

\subsection{Comparison with Analytic Models for Nonlinear Limb Darkening}
I explored using an analytic model to calculate transit light curves for the four-parameter nonlinear limb darkening profile.  The analytic solution for nonlinear limb darkening was presented in \cite{mandel02}, but it is not used in any published software packages. The original code provided by \cite{mandel02} uses a numeric integration scheme that is 20 times slower than the algorithm presented in \S\,\ref{section:algorithm} (for an error tolerance of 1 ppm). 

The analytic solution is challenging to compute because it uses the Appell F1 hypergeometric function.  This function is only convergent for certain regions of parameter space and must be calculated with analytic continuation for other cases. \cite{colavecchia04} provide a Fortran library for computing Appell F1.  I used this library to implement the analytic model for nonlinear limb darkening. However, the returned Appell F1 values are not accurate for all input parameters, based on a comparison with \texttt{Mathematica} and the pure Python library \texttt{mpmath}.  Even for cases where F1 is correct, the computation is over an order of magnitude slower than numeric integration (for an error tolerance of 0.1 ppm).  I concluded that integration is a faster and easier solution than analytic models for the four-parameter nonlinear limb darkening law.

\section{Summary}
\label{section:summary}
I introduced a new algorithm for computing transit light curves for any radially symmetric stellar limb darkening law.  I also described the open-source Python package \texttt{batman}, a versatile code to generate model transit and eclipse light curves. Uniform, linear, quadratic, logarithmic, exponential, and four-parameter nonlinear limb darkening laws are currently supported.  \texttt{batman} uses C extension modules to compute light curves and is parallelized with OpenMP to optimize performance.  Light curves can be calculated with accuracy better than 0.001 ppm, ensuring that the community is prepared to model the extraordinarily precise data we anticipate from upcoming facilities. \texttt{batman} is available at \texttt{\blue{\url{https://github.com/lkreidberg/batman}}} and is also hosted on the Python Package Index under the name \texttt{batman-package}.

\acknowledgments
I thank Jacob Bean, Kevin Stevenson, Eric Agol, Ethan Kruse, Geert Jan Talens, Thomas Beatty, Brett Morris, and Karl Fogel for their support in developing \texttt{batman}. I also thank contributors to SciPy, Matplotlib, and the Python Programming Language for software and services.  I am grateful for helpful suggestions from the referee, Jason Eastman, that improved the code and the manuscript. Support for this work was provided by a grant from the National Science Foundation through a Graduate Research Fellowship to the author.  

\bibliographystyle{apj}
\bibliography{ms.bib}

\begin{thebibliography}{}
\expandafter\ifx\csname natexlab\endcsname\relax\def\natexlab#1{#1}\fi

\bibitem[{{Abubekerov} \& {Gostev}(2013)}]{abubekerov13}
{Abubekerov}, M.~K., \& {Gostev}, N.~Y. 2013, \mnras, 432, 2216

\bibitem[{{Borucki} {et~al.}(2011){Borucki}, {Koch}, {Basri}, {Batalha},
  {Brown}, {Bryson}, {Caldwell}, {Christensen-Dalsgaard}, {Cochran}, {DeVore},
  {Dunham}, {Gautier}, {Geary}, {Gilliland}, {Gould}, {Howell}, {Jenkins},
  {Latham}, {Lissauer}, {Marcy}, {Rowe}, {Sasselov}, {Boss}, {Charbonneau},
  {Ciardi}, {Doyle}, {Dupree}, {Ford}, {Fortney}, {Holman}, {Seager},
  {Steffen}, {Tarter}, {Welsh}, {Allen}, {Buchhave}, {Christiansen}, {Clarke},
  {Das}, {D{\'e}sert}, {Endl}, {Fabrycky}, {Fressin}, {Haas}, {Horch},
  {Howard}, {Isaacson}, {Kjeldsen}, {Kolodziejczak}, {Kulesa}, {Li}, {Lucas},
  {Machalek}, {McCarthy}, {MacQueen}, {Meibom}, {Miquel}, {Prsa}, {Quinn},
  {Quintana}, {Ragozzine}, {Sherry}, {Shporer}, {Tenenbaum}, {Torres},
  {Twicken}, {Van Cleve}, {Walkowicz}, {Witteborn}, \& {Still}}]{borucki11}
{Borucki}, W.~J., {Koch}, D.~G., {Basri}, G., {et~al.} 2011, \apj, 736, 19

\bibitem[{Bulirsch(1965)}]{bulirsch65}
Bulirsch, R. 1965, Numerische Mathematik, 7, 353

\bibitem[{{Charbonneau} {et~al.}(2002){Charbonneau}, {Brown}, {Noyes}, \&
  {Gilliland}}]{charbonneau02}
{Charbonneau}, D., {Brown}, T.~M., {Noyes}, R.~W., \& {Gilliland}, R.~L. 2002,
  \apj, 568, 377

\bibitem[{{Claret}(2000)}]{claret00}
{Claret}, A. 2000, \aap, 363, 1081

\bibitem[{{Claret} \& {Hauschildt}(2003)}]{claret03}
{Claret}, A., \& {Hauschildt}, P.~H. 2003, \aap, 412, 241

\bibitem[{{Colavecchia} \& {Gasaneo}(2004)}]{colavecchia04}
{Colavecchia}, F.~D., \& {Gasaneo}, G. 2004, Computer Physics Communications,
  157, 32

\bibitem[{{Deming} {et~al.}(2013){Deming}, {Wilkins}, {McCullough}, {Burrows},
  {Fortney}, {Agol}, {Dobbs-Dixon}, {Madhusudhan}, {Crouzet}, {Desert},
  {Gilliland}, {Haynes}, {Knutson}, {Line}, {Magic}, {Mandell}, {Ranjan},
  {Charbonneau}, {Clampin}, {Seager}, \& {Showman}}]{deming13}
{Deming}, D., {Wilkins}, A., {McCullough}, P., {et~al.} 2013, \apj, 774, 95

\bibitem[{{Diaz-Cordoves} \& {Gimenez}(1992)}]{diaz-cordoves92}
{Diaz-Cordoves}, J., \& {Gimenez}, A. 1992, \aap, 259, 227

\bibitem[{{Dong} \& {Zhu}(2013)}]{dong13}
{Dong}, S., \& {Zhu}, Z. 2013, \apj, 778, 53

\bibitem[{{Dressing} \& {Charbonneau}(2013)}]{dressing13}
{Dressing}, C.~D., \& {Charbonneau}, D. 2013, \apj, 767, 95

\bibitem[{{Dressing} \& {Charbonneau}(2015)}]{dressing15}
---. 2015, \apj, 807, 45

\bibitem[{{Eastman} {et~al.}(2013){Eastman}, {Gaudi}, \& {Agol}}]{eastman13}
{Eastman}, J., {Gaudi}, B.~S., \& {Agol}, E. 2013, \pasp, 125, 83

\bibitem[{{Foreman-Mackey} {et~al.}(2014){Foreman-Mackey}, {Hogg}, \&
  {Morton}}]{foremanmackey14}
{Foreman-Mackey}, D., {Hogg}, D.~W., \& {Morton}, T.~D. 2014, \apj, 795, 64

\bibitem[{{Fraine} {et~al.}(2014){Fraine}, {Deming}, {Benneke}, {Knutson},
  {Jord{\'a}n}, {Espinoza}, {Madhusudhan}, {Wilkins}, \& {Todorov}}]{fraine14}
{Fraine}, J., {Deming}, D., {Benneke}, B., {et~al.} 2014, \nat, 513, 526

\bibitem[{{Fressin} {et~al.}(2013){Fressin}, {Torres}, {Charbonneau}, {Bryson},
  {Christiansen}, {Dressing}, {Jenkins}, {Walkowicz}, \& {Batalha}}]{fressin13}
{Fressin}, F., {Torres}, G., {Charbonneau}, D., {et~al.} 2013, \apj, 766, 81

\bibitem[{{Gazak} {et~al.}(2012){Gazak}, {Johnson}, {Tonry}, {Dragomir},
  {Eastman}, {Mann}, \& {Agol}}]{gazak12}
{Gazak}, J.~Z., {Johnson}, J.~A., {Tonry}, J., {et~al.} 2012, Advances in
  Astronomy, 2012, 30

\bibitem[{{Gim{\'e}nez}(2006)}]{gimenez06}
{Gim{\'e}nez}, A. 2006, \aap, 450, 1231

\bibitem[{{Howard} {et~al.}(2012){Howard}, {Marcy}, {Bryson}, {Jenkins},
  {Rowe}, {Batalha}, {Borucki}, {Koch}, {Dunham}, {Gautier}, {Van Cleve},
  {Cochran}, {Latham}, {Lissauer}, {Torres}, {Brown}, {Gilliland}, {Buchhave},
  {Caldwell}, {Christensen-Dalsgaard}, {Ciardi}, {Fressin}, {Haas}, {Howell},
  {Kjeldsen}, {Seager}, {Rogers}, {Sasselov}, {Steffen}, {Basri},
  {Charbonneau}, {Christiansen}, {Clarke}, {Dupree}, {Fabrycky}, {Fischer},
  {Ford}, {Fortney}, {Tarter}, {Girouard}, {Holman}, {Johnson}, {Klaus},
  {Machalek}, {Moorhead}, {Morehead}, {Ragozzine}, {Tenenbaum}, {Twicken},
  {Quinn}, {Isaacson}, {Shporer}, {Lucas}, {Walkowicz}, {Welsh}, {Boss},
  {Devore}, {Gould}, {Smith}, {Morris}, {Prsa}, {Morton}, {Still}, {Thompson},
  {Mullally}, {Endl}, \& {MacQueen}}]{howard12}
{Howard}, A.~W., {Marcy}, G.~W., {Bryson}, S.~T., {et~al.} 2012, \apjs, 201, 15

\bibitem[{{Kipping}(2011)}]{kipping11}
{Kipping}, D.~M. 2011, \mnras, 416, 689

\bibitem[{{Kjurkchieva} {et~al.}(2013){Kjurkchieva}, {Dimitrov}, {Vladev}, \&
  {Yotov}}]{kjurkchieva13}
{Kjurkchieva}, D., {Dimitrov}, D., {Vladev}, A., \& {Yotov}, V. 2013, \mnras,
  431, 3654

\bibitem[{{Klinglesmith} \& {Sobieski}(1970)}]{klinglesmith70}
{Klinglesmith}, D.~A., \& {Sobieski}, S. 1970, \aj, 75, 175

\bibitem[{{Knutson} {et~al.}(2014{\natexlab{a}}){Knutson}, {Benneke}, {Deming},
  \& {Homeier}}]{knutson14a}
{Knutson}, H.~A., {Benneke}, B., {Deming}, D., \& {Homeier}, D.
  2014{\natexlab{a}}, \nat, 505, 66

\bibitem[{{Knutson} {et~al.}(2014{\natexlab{b}}){Knutson}, {Dragomir},
  {Kreidberg}, {Kempton}, {McCullough}, {Fortney}, {Bean}, {Gillon}, {Homeier},
  \& {Howard}}]{knutson14b}
{Knutson}, H.~A., {Dragomir}, D., {Kreidberg}, L., {et~al.} 2014{\natexlab{b}},
  \apj, 794, 155

\bibitem[{{Kopal}(1950)}]{kopal50}
{Kopal}, Z. 1950, Harvard College Observatory Circular, 454, 1

\bibitem[{{Kopparapu}(2013)}]{kopparapu13}
{Kopparapu}, R.~K. 2013, \apjl, 767, L8

\bibitem[{{Kreidberg} {et~al.}(2014{\natexlab{a}}){Kreidberg}, {Bean},
  {D{\'e}sert}, {Line}, {Fortney}, {Madhusudhan}, {Stevenson}, {Showman},
  {Charbonneau}, {McCullough}, {Seager}, {Burrows}, {Henry}, {Williamson},
  {Kataria}, \& {Homeier}}]{kreidberg14b}
{Kreidberg}, L., {Bean}, J.~L., {D{\'e}sert}, J.-M., {et~al.}
  2014{\natexlab{a}}, \apjl, 793, L27

\bibitem[{{Kreidberg} {et~al.}(2014{\natexlab{b}}){Kreidberg}, {Bean},
  {D{\'e}sert}, {Benneke}, {Deming}, {Stevenson}, {Seager}, {Berta-Thompson},
  {Seifahrt}, \& {Homeier}}]{kreidberg14a}
---. 2014{\natexlab{b}}, \nat, 505, 69

\bibitem[{{Kreidberg} {et~al.}(2015){Kreidberg}, {Line}, {Bean}, {Stevenson},
  {Desert}, {Madhusudhan}, {Fortney}, {Barstow}, {Henry}, {Williamson}, \&
  {Showman}}]{kreidberg15}
{Kreidberg}, L., {Line}, M.~R., {Bean}, J.~L., {et~al.} 2015, ArXiv e-prints,
  arXiv:1504.05586

\bibitem[{{Lecavelier Des Etangs} {et~al.}(2008){Lecavelier Des Etangs},
  {Pont}, {Vidal-Madjar}, \& {Sing}}]{lecavelier08}
{Lecavelier Des Etangs}, A., {Pont}, F., {Vidal-Madjar}, A., \& {Sing}, D.
  2008, \aap, 481, L83

\bibitem[{{Mandel} \& {Agol}(2002)}]{mandel02}
{Mandel}, K., \& {Agol}, E. 2002, \apjl, 580, L171

\bibitem[{{Morton} \& {Swift}(2014)}]{morton14}
{Morton}, T.~D., \& {Swift}, J. 2014, \apj, 791, 10

\bibitem[{{P{\'a}l}(2012)}]{pal12}
{P{\'a}l}, A. 2012, \mnras, 420, 1630

\bibitem[{{Parviainen}(2015)}]{parvianinen15}
{Parviainen}, H. 2015, \mnras, 450, 3233

\bibitem[{{Petigura} {et~al.}(2013){Petigura}, {Howard}, \&
  {Marcy}}]{petigura13}
{Petigura}, E.~A., {Howard}, A.~W., \& {Marcy}, G.~W. 2013, Proceedings of the
  National Academy of Science, 110, 19273

\bibitem[{{Press} {et~al.}(1992){Press}, {Teukolsky}, {Vetterling}, \&
  {Flannery}}]{press92}
{Press}, W.~H., {Teukolsky}, S.~A., {Vetterling}, W.~T., \& {Flannery}, B.~P.
  1992, {Numerical recipes in FORTRAN. The art of scientific computing}

\bibitem[{{Rowe} {et~al.}(2015){Rowe}, {Coughlin}, {Antoci}, {Barclay},
  {Batalha}, {Borucki}, {Burke}, {Bryson}, {Caldwell}, {Campbell},
  {Catanzarite}, {Christiansen}, {Cochran}, {Gilliland}, {Girouard}, {Haas},
  {He{\l}miniak}, {Henze}, {Hoffman}, {Howell}, {Huber}, {Hunter},
  {Jang-Condell}, {Jenkins}, {Klaus}, {Latham}, {Li}, {Lissauer}, {McCauliff},
  {Morris}, {Mullally}, {Ofir}, {Quarles}, {Quintana}, {Sabale}, {Seader},
  {Shporer}, {Smith}, {Steffen}, {Still}, {Tenenbaum}, {Thompson}, {Twicken},
  {Van Laerhoven}, {Wolfgang}, \& {Zamudio}}]{rowe15}
{Rowe}, J.~F., {Coughlin}, J.~L., {Antoci}, V., {et~al.} 2015, \apjs, 217, 16

\bibitem[{{Schwarzschild} \& {Villiger}(1906)}]{schwarzschild06}
{Schwarzschild}, K., \& {Villiger}, W. 1906, \apj, 23, 284

\bibitem[{{Seager}(2010)}]{seager10}
{Seager}, S. 2010, {Exoplanets}

\bibitem[{{Seager} \& {Sasselov}(2000)}]{seager00}
{Seager}, S., \& {Sasselov}, D.~D. 2000, \apj, 537, 916

\bibitem[{{Sing} {et~al.}(2011){Sing}, {Pont}, {Aigrain}, {Charbonneau},
  {D{\'e}sert}, {Gibson}, {Gilliland}, {Hayek}, {Henry}, {Knutson}, {Lecavelier
  Des Etangs}, {Mazeh}, \& {Shporer}}]{sing11}
{Sing}, D.~K., {Pont}, F., {Aigrain}, S., {et~al.} 2011, \mnras, 416, 1443

\bibitem[{{Southworth} {et~al.}(2004){Southworth}, {Maxted}, \&
  {Smalley}}]{southworth04}
{Southworth}, J., {Maxted}, P.~F.~L., \& {Smalley}, B. 2004, \mnras, 351, 1277

\bibitem[{{Stevenson} {et~al.}(2014{\natexlab{a}}){Stevenson}, {Bean},
  {Fabrycky}, \& {Kreidberg}}]{stevenson14c}
{Stevenson}, K.~B., {Bean}, J.~L., {Fabrycky}, D., \& {Kreidberg}, L.
  2014{\natexlab{a}}, \apj, 796, 32

\bibitem[{{Stevenson} {et~al.}(2014{\natexlab{b}}){Stevenson}, {Bean},
  {Madhusudhan}, \& {Harrington}}]{stevenson14b}
{Stevenson}, K.~B., {Bean}, J.~L., {Madhusudhan}, N., \& {Harrington}, J.
  2014{\natexlab{b}}, ArXiv e-prints, arXiv:1406.7567

\bibitem[{{Stevenson} {et~al.}(2014{\natexlab{c}}){Stevenson}, {Bean},
  {Seifahrt}, {D{\'e}sert}, {Madhusudhan}, {Bergmann}, {Kreidberg}, \&
  {Homeier}}]{stevenson14a}
{Stevenson}, K.~B., {Bean}, J.~L., {Seifahrt}, A., {et~al.} 2014{\natexlab{c}},
  \aj, 147, 161

\bibitem[{{Stevenson} {et~al.}(2014{\natexlab{d}}){Stevenson}, {D{\'e}sert},
  {Line}, {Bean}, {Fortney}, {Showman}, {Kataria}, {Kreidberg}, {McCullough},
  {Henry}, {Charbonneau}, {Burrows}, {Seager}, {Madhusudhan}, {Williamson}, \&
  {Homeier}}]{stevenson14d}
{Stevenson}, K.~B., {D{\'e}sert}, J.-M., {Line}, M.~R., {et~al.}
  2014{\natexlab{d}}, Science, 346, 838

\bibitem[{{Traub}(2012)}]{traub12}
{Traub}, W.~A. 2012, \apj, 745, 20

\bibitem[{{Youdin}(2011)}]{youdin11}
{Youdin}, A.~N. 2011, \apj, 742, 38

\end{thebibliography}

\end{document}